\documentclass[aps,prl,groupedaddress,]{revtex4}
\usepackage{latexsym}
\usepackage[dvips]{graphicx}

\newcommand{\eq}{\begin{equation}}
\newcommand{\feq}{\end{equation}}
\newcommand{\eqn}{\begin{eqnarray}}
\newcommand{\feqn}{\end{eqnarray}}
\newcommand{\arr}{\begin{eqnarray*}}
\newcommand{\farr}{\end{eqnarray*}}
\newcommand{\beq}{\begin{equation}}
\newcommand{\eeq}{\end{equation}}
\newcommand{\bea}{\begin{eqnarray}}
\newcommand{\eea}{\end{eqnarray}}

\def\beq{\begin{equation}}
\def\eeq{\end{equation}}
\def\feq{\end{equation}}
\def\bea{\begin{eqnarray}}
\def\eea{\end{eqnarray}}
\def\bc{\begin{displaymath}}
\def\ec{\end{displaymath}}
\def\lb{\label}

\def\xp{{\xi^\parallel}}

\def\la{\lambda}

\def\ep{\varepsilon}

\def\xm{{x^{-}}}
\def\xp{x^{+}}
\def\lxm{{\lambda x^{-}}}

\def\lb{\label}
\def\nn{\nonumber}
\def\ord#1{O\left(#1\right)}

\begin{document}


\title{Statistical Entropy of the Schwarzschild black hole}

\author{Mariano Cadoni}
\email{mariano.cadoni@ca.infn.it}
\affiliation{Dipartimento di Fisica,
Universit\`a di Cagliari, and INFN sezione di Cagliari, Cittadella
Universitaria 09042 Monserrato, ITALY}


\begin{abstract}
We derive the statistical entropy of the Schwarzschild black hole by considering 
the asymptotic symmetry algebra near the $\cal{I^{-}}$  
boundary of the spacetime at past null infinity. Using a two-dimensional description  
and the Weyl invariance  of 
black hole thermodynamics this symmetry algebra can be mapped into the 
Virasoro algebra 
generating  asymptotic symmetries of  anti-de Sitter spacetime.
Using lagrangian methods we identify the stress-energy tensor 
of the boundary conformal field theory and  we calculate the  central charge 
of the Virasoro algebra. The Bekenstein-Hawking  result for the black 
hole entropy is regained using  Cardy's formula. Our result strongly 
supports a non-local realization of the holographic principle 
\end{abstract}


\maketitle
Black holes can be understood as thermodynamical systems with 
characteristic temperature and entropy \cite{Bekenstein:1972tm,
Hawking:1974sw}. In the last decade 
a lot of effort  has been devoted  to understand the 
microscopic origin of black hole 
thermodynamics.  A detailed microscopical explanation of the 
thermodynamical properties of black holes  would represent not only an
important tool for understanding  the quantum behavior of gravity but 
also a way to give a fundamental meaning to the holographic principle 
\cite{Susskind:1994vu,Bousso:2002ju}.
A variety of approaches have been used to explain the microscopical 
origin of the Bekenstein-Hawking entropy: string 
theoretical  (D-Brane)  approaches 
\cite{Strominger:1996sh,Horowitz:1996fn,Johnson:1996ga,Callan:1996dv,
Aharony:1999ti}, methods based on 
loop quantum gravity \cite{Ashtekar:1997yu}, induced gravity 
\cite{Frolov:1997up,Fursaev:2004qz}, 
asymptotic  symmetries 
\cite{Strominger:1997eq,Cadoni:1998sg,Cadoni:1999ja,Carlip:1998wz,
 Caldarelli:2000xk,Navarro:1999qy,Jing:2000yn,Park:2001zn,
Carlip:2002be,Medved:2002dw,Carlip:2004mn}
and canonical quantization 
\cite{Moretti:2005jz}.
None of these derivations can be  considered as completely 
satisfactory. In some cases the computation works only for a restrict class of 
solutions, e.g supersymmetric or asymptotically Anti-de Sitter (AdS) 
solutions. In other cases the origin of the microscopic degrees of 
freedom responsible for the black hole entropy is completely obscure.
Nonetheless, the fact that completely independent methods give more or 
less the same (right) answer  strongly indicates that the 
computation of the statistical black hole entropy can be performed 
without detailed knowledge of the physics governing the microscopical 
degrees of freedom.

If this is the case the statistical black hole entropy  should be
explained in terms of fundamental features (e.g   symmetries) of the 
``emergent'' classical theory of gravity of which black holes are 
solutions. The  most natural realization of this 
scenario   is  represented by  near-horizon conformal symmetries controlling the 
black hole entropy trough Cardy's formula for the density of states 
\cite{Carlip:2002be,Carlip:2004mn}.
Carlip has pointed out the main ingredients to be used in this approach:
$a)$ Near-horizon 
conformal symmetries, $b)$  Asymptotic symmetries,  $c)$ Canonical 
realization of the symmetries, $d)$ Horizon constraints \cite{Carlip:2002be}.
However, the approach of Carlip has some drawbacks. To avoid the 
complications of the  Hamiltonian
formulation  with a null slicing 
one has to 
consider  a distorted, ``almost'' null, horizon \cite{Carlip:2002be,Carlip:2004mn}. 
The null limit of this formulation is 
usually singular and the null normal does not have  a unique normalization.
Moreover, the  result seems to depend on the horizon boundary 
conditions  
\cite{Pinamonti:2003bh,Kang:2004js} and the constraint algebra is complicated by the appearance 
of additional constraints related to the use of a null frame in the 
slicing \cite{Carlip:2004mn}.

Apart from these technical  difficulties,  derivations that use  
near-horizon symmetries leave unanswered the question about the 
localization of the microstates responsible for the black hole entropy.
A number of cases are known for which the Bekenstein-Hawking entropy 
can be reproduced using a conformal filed theory (CFT)
living either on the black hole 
horizon or on a timelike asymptotical boundary of the spacetime 
\cite{Strominger:1997eq,Cadoni:1998sg,Cadoni:1999ja,Carlip:1998wz,
Caldarelli:2000xk}.
An answer to this question is of fundamental 
importance for understanding if the holographic principle is 
realized in a local or non-local way.

In this letter we will avoid the subtleties of the canonical approach 
by considering a purely lagrangian formulation of the gravity theory. 
We will derive the 
statistical entropy of the Schwarzschild black hole by considering 
the asymptotic symmetry algebra near the $\cal{I^{-}}$   
 boundary of the spacetime at past null infinity. Using a two-dimensional (2D) model to describe 
the radial modes of Einstein gravity and the Weyl invariance  of 
black hole thermodynamics \cite{Cadoni:1996bn}, this symmetry algebra 
is mapped in the 
Virasoro algebra 
generating  asymptotic symmetries of 2D  AdS spacetime.
Using lagrangian methods, we identify the 
stress-energy tensor for  the boundary conformal field theory 
and we calculate the   central charge of the Virasoro algebra.
The Bekenstein-Hawking  result for the black 
hole entropy is obtained from  Cardy's formula.  

The Schwarzschild black hole of general relativity  behaves as a 
thermodynamical system with temperature $T$ and entropy $S$  given by  
\beq\lb{bhp}
T=\frac{1}{8\pi G M},\quad \, S=4\pi G M^{2},
\feq
where $M$ is the black hole mass and $G$ is Newton constant.
The  black hole   admits   
a two-dimensional  effective description. The  effective 2D gravity model 
can be obtained from the four-dimensional (4D)  Einstein action retaining only the radial 
modes of the gravitational field,
\beq\lb{dm}
ds_{(4)}=ds_{(2)}+ \frac{2}{\la^{2}}\phi d\Omega^{2}_{2},
\feq
where $\la=1/\sqrt{G}$  is the Planck mass (we use natural units) 
and $\phi$ is a scalar field parametrizing the 
radius of the transverse 2-Sphere. Because the causal structure and 
the  thermodynamical parameters (mass $M$, temperature $T$ and 
entropy $S$)  of the 2D solutions are invariant 
under Weyl rescaling of the   2D metric \cite{Cadoni:1996bn},  
the Schwarzschild solution 
can be described by an equivalence class (under conformal transformations 
of the metric) of 2D gravity models. 
Let us use this  freedom to pick up a 2D gravity model with 
asymptotic anti-de Sitter behavior.  

Performing the dimensional 
reduction (\ref{dm}) in the 4D Einstein-Hilbert action and the Weyl 
rescaling of the 2D metric 
$g^{(2)}¥_{\mu\nu}\to(2\phi)g^{(2)}¥_{\mu\nu}$, 
we end up with the 2D dilaton gravity model,
\beq\lb{2dm}
S=\frac{1}{2}\int d^{2}x \sqrt{-g}\left(\phi R +\frac{3}{2} \frac{(\nabla 
\phi)^{2}}{\phi}+2\la^{2} \phi\right).
\feq
The static solutions of the gravity model (\ref{2dm}) are 
\beq\lb{2dsol}
ds^{2}=-\left[(\la x)^{2}- \frac{2M}{\la} (\la x)^{3}\right]dt^{2}+
\left[(\la x)^{2}- \frac{2M}{\la} (\la 
x)^{3}\right]^{-1}¥dx^{2},\quad \phi= \frac{1}{2}(\la x)^{-2}.
\feq
One can easily check that the solution (\ref{2dsol}) describes the 2D 
sections of the 4D Schwarzschild black hole after performing the Weyl 
rescaling $\hat g_{\mu\nu}=(2\phi)g_{\mu\nu}$  and changing the 
radial coordinate $\la x=1/(\la r).$
Using well-known formulas, one can  easily  calculate the 2D 
thermodynamical parameters associated with  solution (\ref{2dsol}),  
reproducing exactly those of the Schwarzschild black hole (\ref{bhp}).

Solution (\ref{2dsol}) is asymptotically AdS
but not in the usual 
sense of its $x\to \infty$ behavior (For a discussion  of 
2D AdS spacetime (AdS$_{2}$) and its different parametrization see 
Ref. \cite{Cadoni:1994uf})). It behaves as AdS$_{2}$ for 
$x\to 0$, when it takes the form $ds^{2}=-(\la x)^{2}dt^{2}+(\la 
x)^{-2}dx^{2}$. The exterior region of the Schwarzschild black hole, $ 
2MG<r<\infty$, is mapped by the a Weyl transformation and 
 coordinate transformation  
into the region  $0<x<1/(2M)$ of the 2D solution (\ref{2dsol}). 
The corresponding Penrose diagrams of the solutions are also the same 
in the considered range of the coordinates $r, x$. 
They have both  
the shape of a diamond, 
with the ${\cal I} ^{\pm}$, $r\to\infty$ null 
boundaries (past and future horizon ${\cal H}^{\pm}$ at $r=2GM$) of the 
Schwarzschild spacetime mapped into the  internal $I^{\pm}$, 
$x=0$ null boundaries (past and future horizon $ H^{\pm}$ at 
$x=1/(2M)$) of the spacetime (\ref{2dsol}).

To discuss the behavior  of the 2D gravity model  on the null 
boundaries $I^{\pm}$ we need to write the solution (\ref{2dsol}) 
using light-cone coordinates,
\beq\lb{lc}
ds^{2}=2g_{+-}dx^{+}dx^{-}= -\left((\la x)^{2}-\frac{2M}{\la}(\la 
x)^{3}\right)dx^{+}dx^{-}, \quad \phi=\frac{1}{2}(\la x)^{-2},
\feq
where $x^{\pm}$ are implicitly defined by  
$(\la/2)(x^{+}-x^{-})= (\la x)^{-1}+ (2M/\la)[\ln(1-2M 
x)-\ln (\la x)], \quad t=(1/2)(x^{+}+x^{-})$.

The  symmetry relevant for our discussion is the group of the asymptotic 
symmetries (ASG) of  the Schwarzschild spacetime near ${\cal I} ^{\pm}$. 
A  discussion of the ASG for asymptotically flat 
spacetimes can be found in  Ref. \cite{Dappiaggi:2004kv}. The previously 
discussed Weyl 
transformation maps the ${\cal I} ^{\pm}$  boundary of the 
Schwarzschild spacetime into the $I^{\pm}$ inner boundary of our 2D 
spacetime, hence  we can equivalently consider the ASG of    
our model (\ref{2dm}) for $x\to 0$. Our 2D spacetime behaves 
for $x\to 0$ as AdS$_{2}$, we therefore expect the ASG to be the conformal 
group in one dimension. 

The $x=0$ inner boundary of the spacetime splits 
in the two orthogonal parts $I^{+}$ and $I^{-}$.  When discussing the 
ASG we are  forced to consider only one of those two parts. We will 
choose $I^{-}$, that is we discuss the asymptotic behavior of the 
solutions for $x^{-}\to -\infty$ and $0< x^{+}<\infty$.
For $x^{-}\to -\infty$ solution (\ref{lc}) behaves as
\beq\lb{ab}
ds^{2}= -\left[ \frac{4}{{(\lxm)}^{2}} +\ord 
{\frac{\ln(-\lxm)}{{(\xm)}^{3}}}\right]dx^{+}dx^{-}, \quad 
\phi=\frac{{(\lxm)}^{2}}{8}+\ord {\xm\ln(-\lxm)}.
\feq
We are therefore led to impose the following boundary conditions at 
$\xm\to-\infty$
\bea\lb{bc}
g_{+-}&=& - \frac{2}{{(\lxm)}^{2}} +\ord 
{\frac{\ln(-\lxm)}{{(\xm)}^{3}}}, \quad g_{++}=\ord 
{\frac{1}{\xm}},\nn\\
g_{--}&=& 0\,\,\, ({\rm identically}),\quad 
\phi=\frac{{(\lxm)}^{2}}{8}+\ord {\xm\ln(-\lxm)}.
\eea
The asymptotic form (\ref{bc}) is preserved by infinitesimal 
diffeomorphisms $\chi^{\mu}(\xm,\xp)$ given by
\beq\lb{id}
\chi^{+}=\ep(\xp),\quad
\chi^{-}= \dot\ep(\xp) \xm + \ord 1,
\feq
where the dot denotes derivation 
with respect to $\xp$.
Expanding $\ep(\xp)$ in Laurent series yields the generators of 
the ASG,
\beq\lb{ge}
L_{k}=\left[(k+1)\xm(\la\xp)^{k}+\ord1
\right]\partial_{-}+\frac{1}{\la}\left[(\la\xp)^{k+1}+\ord{\frac{1}{\xm}}
\right]\partial_{+},
\feq
which satisfy (after the sign flipping $L_{k} \to - L_{k}$) the 
Virasoro algebra
\beq\lb{va}
[L_{k},L_{m}]= (k-m)L_{k+m}+ \frac{c}{12}k^{3}\delta_{k+m\, 0},
\feq
where we have taken into account the possibility of a central 
extension $c$.

The boundary conditions (\ref{bc}) allows us to define boundary 
fields $\Gamma_{0}¥(\xp),\Gamma_{1}(\xp)\ldots,$
$\hat\Gamma_{0}¥(\xp),\hat\Gamma_{1}(\xp)\ldots,$  $\gamma_{0}(\xp),
\gamma_{1}(\xp),\ldots$, which describe deformations of the 
boundary $I^{-}$ and of the scalar field $\phi$,
\bea\lb{def}
g_{+-}&=& - \frac{1}{2}\left[\frac {1}{{X}^{2}} + 
\Gamma_{0}\frac{\ln(-X)}{{X}^{3}}+\frac{\Gamma_{1}}{{X}^{3}}+
\Gamma_{2}\frac{\ln^{2}¥(-X)}{{X}^{4}}+\ldots\right],\nn\\
g_{++}&=& 
\frac{\hat\Gamma_{0}}{X}+
\hat\Gamma_{1}\frac{\ln¥(-X)}{{X}^{2}}+\ldots,\\
\phi&=&  \gamma_{0}¥X^{2}¥ +\gamma_{1}¥X \ln(-X)+ 
\gamma_{2}¥X + \gamma_{3} \ln^{2}¥(-X)+ \gamma_{4}\ln(-X)+
\gamma_{5}+\ldots,\nn\\\nn
\eea
where $X= (\la \xm/2)$.

It is important to notice that the leading terms  for the metric and  
the scalar $\phi$ in the expansion 
(\ref{def}) ( hence also in Eq. (\ref{bc})) are not  solution of the 
classical field equation coming 
from the action (\ref{2dm}). The classical black hole solutions can be recovered 
only choosing a particular form of the   boundary fields in Eqs. (\ref{def}).
This means that we are using a notion of  asymptotic symmetry which is 
slightly different form the usual one (see Ref. \cite{Brown:1986nw}).
In the usual formulation the leading term in the boundary conditions 
represents a (background) classical solution that remains invariant 
under the action of the ASG. In our formulation what is invariant 
under transformation of the ASG is the leading term in the asymptotic form of 
the spacetime metric, which does not need to be a solution of the 
field  equations.

The boundary fields $\Gamma_{i},\hat\Gamma_{i},\gamma_{i}, \, i=0,1\ldots$ 
transform 
under the action of the ASG as  conformal fields of definite weight 
( with possible anomalous terms). 
We have for instance $\delta \Gamma_{0}= \ep 
\dot\Gamma_{0}-\Gamma_{0}\dot \ep,$ $\delta \gamma_{0}= \ep 
\dot\gamma_{0}+2\Gamma_{0}\dot \ep,$.  In the following we will need  the 
transformation law for the boundary field $\gamma_{5}$,
\beq\lb{g5}
\delta\gamma_{5}= \ep \dot\gamma_{5}+\dot \ep\gamma_{4}.
\feq

The next step in our derivation is to define the  charges 
associated  with the generators of the ASG $L_{k}$.  To define the 
charges one usually considers a canonical realization of the ASG 
\cite{Brown:1986nw}. The Hamiltonian approach works well when dealing 
with timelike  (also spacelike, see Ref. \cite{Cadoni:2002kz}) boundaries, is problematic  
for null boundaries \cite{Carlip:2002be,Carlip:2004mn}.  In fact, foliating the 2D spacetime using 
null curves one has  ambiguities connected with the use of null 
vectors. On the other hand, if one uses a distorted
boundary and a corresponding foliation  with spacelike curves, one 
usually has a singular limit when the boundary becomes null. To avoid 
these problems we will use  a purely  Lagrangian method to define the charges 
associated with the ASG generators.

The covariantly conserved current  $J_{\mu}$ associated with   an isometry of 
the  spacetime generated by a Killing vector $\chi^{\nu}$ can be 
 written has  $J_{\mu}=T_{\mu\nu}\chi^{\nu}$, where 
$T_{\mu\nu}=(2/\sqrt{-g})(\delta S/\delta g^{\mu\nu})$ is the 
stress-energy tensor. In two spacetime 
dimensions one can always integrate, locally, the equation 
$\nabla^{\mu}J_{\mu}=0$, yielding $J_{\mu}={\bf 
\epsilon}^{\nu}_{\mu}\partial _{\nu} \Omega$, where $\Omega$ is a scalar, 
which  can be considered as
the charge associated with 
the isometry generated by $\chi^{\nu}$,
\beq\lb{e1}
{\bf 
\epsilon}^{\nu}_{\mu}\partial _{\nu} \Omega=T_{\mu\nu}\chi^{\nu}.
\feq
The classical equations of motion give $T_{\mu\nu}=0,$ so 
that on-shell the charges are constant. 
This definition can be easily generalized to define the charges 
associated with the generators of the group of asymptotic symmetries.
Using light-cone coordinates Eq. (\ref{e1}) yields 
$\partial_{+}\Omega=-(T_{++}\chi^{+}+T_{+-}\chi^{-}),\,
\partial_{-}\Omega=T_{--}\chi^{-}+T_{-+}\chi^{+}$.
Pushing these equations on the $I^{-}$ boundary by taking the $\xm\to 
-\infty$ limit, the second equation becomes an identity, whereas the 
first yields the definition of the charges associated with the ASG 
generated by the killing vectors (\ref{id}),
\beq\lb{charges}
\Omega(\ep)=-\lim_{\xm\to-\infty}\int_{I^{-}¥}d\xp\, 
\left(\ep T_{++}+\dot \ep \xm T_{+-}\right).
\feq
The term proportional to $\xm T_{+-}$ vanishes on the $I^{-}$ 
boundary and we have $\Omega(\ep)=\int d\xp 
\ep T, $ where $T=- \lim_{\xm\to-\infty}T_{++}$.
$T$ has to be considered as the stress-energy tensor for a 
one-dimensional CFT living on $I^{-}$.
The tensor $T$ can be derived  from the action (\ref{2dm}) and using 
Eqs. (\ref{def}). $T$  is identically zero when the boundary fields 
(\ref{def}) are on-shell,  i.e when they describe a classical 
black hole solution of the action (\ref{2dm}). For generic off-shell 
deformations $T$ is the sum of terms, which diverge for $\xm 
\to - \infty$ (the leading divergence is $\propto ({\xm})^{2}$), and  
finite terms. The simplest way to eliminate these divergences is to 
consider all deformations  (\ref{def}) on-shell except $\gamma_{5}$.
Using this renormalization  prescription we get
\beq\lb{st}
T= \ddot\gamma_{5}.
\feq
This procedure has a simple physical interpretation. We start from 
 background  boundary deformations  $\Gamma_{i}^{(b)}¥,\gamma_{i}^{(b)}¥$ 
corresponding to classical black hole solution of a given mass $M$
and allow only for off-shell boundary deformations of the classical 
configuration that produce charges, whichÇÈ  are finite for $\xm\to 
-\infty$.

The transformation law of $T$ under the action of the ASG
can be easily derived using Eq. 
(\ref{g5}). We have
\beq\lb{tl}
\delta T= \ep\dot T+2\dot \ep T+ C(\ep,\gamma_{4}, \gamma_{5}),
\feq
where $C= \gamma_{4}\stackrel{...}{\ep}+\ddot\gamma_{4}\dot \ep+
(\dot\gamma_{5}+ 2\dot\gamma_{4})\ddot\ep$.
As expected Eq. (\ref{tl}) is the transformation law  of 
a stress-energy tensor of a CFT 
living on the $I^{-}$ boundary.
The anomalous term $C$ has to be evaluated 
on the classical background 
configuration, $\gamma_{4}^{(b)}¥= 
(2M/\la)^{2}- M\xp,\, \gamma_{5}^{(b)}¥= 
(2M/\la)^{2}-M\xp + (\la^{2}/8)(\xp)^{2}$. For macroscopic black 
holes $(M/\la)>>1$. Retaining 
only the leading term for $(M/\la)>>1$, $C$ takes its standard form, 
$C=\left(\frac{2M}{\la}\right)^{2}\stackrel{...}{\ep}$. 
From this equation one can read 
off the value of the central charge of the boundary CFT,
\beq\lb{cc}
c= 48 \frac{M^{2}}{\la^{2}}.
\feq

Apart from the divergence for $\xm\to\infty$,  the charges $\Omega$  
are also infrared divergent, owing to the 
non-compactness of the $I^{-}$ boundary. 
As a further consequence of this divergence the inner product on 
$I^{-}$ is ill-defined preventing a realization of the algebra 
(\ref{va})  trough the charges (\ref{charges}). 
Both problems can be solved introducing an infrared cutoff $R$ and 
compactifying the $I^{-}$ boundary.  

The size $R$ of  compactified $I^{-}$ is set by the temperature 
(\ref{bhp}) of the 
Schwarzschild black hole. $R$ has to be 
identified in terms  of the periodicity $\beta=1/T$  of the 
euclidean time: $R= (1/2) (\beta/2\pi)= 2M/\la^{2}$ ( the $1/2$ 
factor takes into account the presence of the orthogonal $I^{+}$ 
boundary).
The  charges are now finite and become,
\beq\lb{rc}
\Omega(\ep)=\int_{0}^{R}d \xp \ep T.
\feq
Using this expression one can easily calculate the charge ( the 
eigenvalue) of the Virasoro operator $L_{0}$.
Reading in Eq. (\ref{ge}) the value of $\ep$ corresponding to the 
$L_{0}$ operator,  ($\ep=\xp$) and using Eq. (\ref{st}) evaluated 
on-shell into Eq. (\ref{rc}), one gets
\beq\lb{l0}
\Omega(L_{0})=\frac{M^{2}}{2\la^{2}}.
\feq

The Virasoro algebra (\ref{va}) and in particular the central charge 
$c$  are determined by  the short distance behavior of the theory 
and are independent of both the infrared cutoff $R$ and the topology 
of the $I^{-}$ boundary. We can therefore give a realization of the 
Virasoro algebra (\ref{va}) in terms of the charges (\ref{rc})
by considering a (infrared regularized) boundary with the topology 
of  $S^{1}$ and writing the transformation law of $\Omega(\ep)$ in 
operatorial form,
 \beq\lb{e5}
\delta_{\xi}\Omega(\ep)=i[\Omega(\ep),\Omega(\xi)],
\feq
where  $\ep,\xi$ parametrize two infinitesimal ASG transformations. 
Expanding in Fourier modes $\Omega(\ep)=\sum_{m}a_{m}L_{m}$, 
$\ep=[R/(2\pi)^{3/2}] \sum_{m}a_{m}f_{m}$,  $T=(2\pi/R)^{2}¥\sum_{m}L_{m}f_{-m},$ 
with 
$f_{m}=\exp(2\pi i m\xp/R)$ and using Eqs. (\ref{tl}), 
(\ref{rc}) into
(\ref{e5}),  one 
finds that the  operators $L_{m}$ satisfy the algebra (\ref{va}) with 
central charge given exactly by Eq. (\ref{cc}).

The entropy associated with the boundary CFT characterized by  
 eigenvalue $l_0$ of the operator $L_0$ and central charge $c$, 
is given in the semiclassical regime we are considering  by the Cardy formula  
$S=2\pi \sqrt{c l_0/6}$.
Using Eqs. (\ref{cc}) and (\ref{l0}) we get for the 
entropy,
\beq\label{Angius}
S=4\pi\frac{M^{2}}{\la^{2}},
\feq
matching exactly the Bekenstein-Hawking entropy of the Schwarzschild 
black hole.

The main result of this paper is that the Bekenstein-Hawking entropy 
of the Schwarzschild black hole can be correctly reproduced by  
counting states of a CFT defined on the $\cal{I^{-}}$  
boundary of the spacetime at past null infinity.  
Because the same result  can be obtained 
using a CFT defined on the black hole horizon, this seems to imply a 
non-local realization of the holographic principle.  
Our derivation method uses a purely lagrangian formulation of the 
gravity theory and it is 
therefore free of the usual  problems   affecting the Hamiltonian 
formulation in  null frames. 
The divergences of the boundary stress-energy tensor for 
$\xm\to-\infty$ can be removed 
in a non-ambiguous way. The only  source of ambiguity comes from 
the infrared divergence due to the infinite length of the $I^{-}$
boundary. This divergence has  been  removed by compactification.
The radius of the compactified boundary has been 
fixed in a natural way in terms of inverse temperature of the black 
hole.

\begin{acknowledgments}
I am grateful to N. Pinamonti for  discussions about 
the ambiguities of the canonical formulation in the presence of null 
boundaries
\end{acknowledgments}

\end{document}